\documentclass[preprint,aps,superscriptaddress,floatfix]{revtex4-1}
\usepackage{graphicx}
\usepackage{amsmath}
\usepackage{natbib}
\usepackage{amsfonts}

\begin{document}

\title{Layer-dependent spin-orbit torques generated by the centrosymmetric transition metal dichalcogenide $\beta$-MoTe$_2$}

\author{Gregory M. Stiehl}
\affiliation{Department of Physics, Cornell University, Ithaca, NY 14853, USA}
\author{Ruofan Li}
\affiliation{Department of Physics, Cornell University, Ithaca, NY 14853, USA}
\author{Vishakha Gupta}
\affiliation{Department of Physics, Cornell University, Ithaca, NY 14853, USA}
\author{Ismail El Baggari}
\affiliation{Department of Physics, Cornell University, Ithaca, NY 14853, USA}

\author{Shengwei Jiang}
\affiliation{School of Applied and Engineering Physics, Cornell University, Ithaca, NY 14853, USA}
\author{Hongchao Xie}
\affiliation{School of Applied and Engineering Physics, Cornell University, Ithaca, NY 14853, USA}
\author{Lena F. Kourkoutis}
\affiliation{School of Applied and Engineering Physics, Cornell University, Ithaca, NY 14853, USA}
\affiliation{Kavli Institute at Cornell, Cornell University, Ithaca, NY 14853, USA}
\author{Kin Fai Mak}
\affiliation{Department of Physics, Cornell University, Ithaca, NY 14853, USA}
\affiliation{School of Applied and Engineering Physics, Cornell University, Ithaca, NY 14853, USA}
\affiliation{Kavli Institute at Cornell, Cornell University, Ithaca, NY 14853, USA}
\author{Jie Shan}
\affiliation{Department of Physics, Cornell University, Ithaca, NY 14853, USA}
\affiliation{School of Applied and Engineering Physics, Cornell University, Ithaca, NY 14853, USA}
\affiliation{Kavli Institute at Cornell, Cornell University, Ithaca, NY 14853, USA}
\author{Robert A. Buhrman}
\affiliation{School of Applied and Engineering Physics, Cornell University, Ithaca, NY 14853, USA}
\author{Daniel C. Ralph}
\affiliation{Department of Physics, Cornell University, Ithaca, NY 14853, USA}
\affiliation{Kavli Institute at Cornell, Cornell University, Ithaca, NY 14853, USA}
 \email{dcr14@cornell.edu}
\date{{\small \today}}
\date{{\small \today}}

\begin{abstract}

Single-crystal materials with sufficiently low crystal symmetry and strong spin-orbit interactions can be used to generate novel forms of spin-orbit torques on adjacent ferromagnets, such as the out-of-plane antidamping torque previously observed in WTe$_2$/ferromagnet heterostructures. Here, we present measurements of spin-orbit torques produced by the low-symmetry material $\beta$-MoTe$_2$, which unlike WTe$_2$ retains bulk inversion symmetry. We measure spin-orbit torques on $\beta$-MoTe$_2$/Permalloy heterostructures using spin-torque ferromagnetic resonance as a function of crystallographic alignment and MoTe$_2$ thickness down to the monolayer limit. We observe an out-of-plane antidamping torque with a spin torque conductivity as strong as 1/3 of that of WTe$_2$, demonstrating that the breaking of bulk inversion symmetry in the spin-generation material is not a necessary requirement for producing an out-of-plane antidamping torque. We also measure an unexpected dependence on the thickness of the $\beta$-MoTe$_2$ -- the out-of-plane antidamping torque is present in MoTe$_2$/Permalloy heterostructures when the $\beta$-MoTe$_2$ is a monolayer or trilayer thick, but goes to zero for devices with bilayer $\beta$-MoTe$_2$. 

\end{abstract}
\maketitle

Spin-orbit torques represent one of the most promising methods for manipulating emerging magnetic memory technologies\ \cite{Brataas2012}. When a charge current is applied to a material with large spin-orbit coupling, such as a heavy metal\ \cite{Miron2010,Miron2011,Liu2011,LiuPRL2012,Liu555,Pai2012}, topological insulator\ \cite{Mellnik2014,Fan2014}, or transition metal dichalcogenide (TMD)\ \cite{ZhangAPL2016,Shao2016,MacNeill2016,MacNeill2017,Guimaraes2018Nano,Li2018,Stiehl2018}, a spin current generated through mechanisms such as the spin Hall or Rashba-Edelstein effects can be used to exert a torque on an adjacent ferromagnet. Recent work from several research groups has focused on understanding how a controlled breaking of symmetry in a spin-generating material / ferromagnet heterostructure can be used to tune the direction of the observed spin-orbit torques for optimal switching of magnetic devices\ \cite{Baek2018,Safranski2018,Gibbons2018,Ou2018,Fang2011,Kurebayashi2014,Skinner2015,Ciccarelli2016,MacNeill2016,MacNeill2017,Guimaraes2018Nano}. For instance, the presence of magnetic order within a spin-generation layer can allow current-generated spin directions that are typically forbidden for highly-symmetric non-magnetic metals\ \cite{Baek2018,Safranski2018,Gibbons2018,Ou2018}. Similarly, our group has shown that by using WTe$_2$ as the spin-source material, a TMD with a low-symmetry crystal structure, it is possible to generate an out-of-plane antidamping torque\ \cite{MacNeill2016,MacNeill2017} -- the component of torque required for the most efficient mode of switching for magnets with perpendicular magnetic anisotropy, but forbidden in higher-symmetry materials. Only one other material, SrRuO$_3$, has been shown to generate an out-of-plane antidamping spin-orbit torque\ \cite{Ou2018}, arising from symmetry breaking associated with magnetic order. Many questions remain regarding the mechanism and necessary conditions for generating a strong out-of-plane antidamping torque. 

In this work, we study the spin-orbit torques generated in TMD/ferromagnet heterostructures with a crystal symmetry that is distinct from WTe$_2$ in an important way -- inversion symmetry is intact in the bulk of the spin-generation material. We perform spin-torque measurements of TMD/ferromagnet heterostructures with the monoclinic phase ($\beta$) of MoTe$_2$ as the spin-source material. Using spin-torque ferromagnetic resonance (ST-FMR), we measure the spin-orbit torques  as a function of crystal axis alignment and TMD thickness down to the monolayer limit. We find that an out-of-plane antidamping torque is present in $\beta$-MoTe$_2$/ferromagnet heterostructures even though inversion symmetry is intact in the MoTe$_2$ bulk. Interestingly, we find that while this out-of-plane antidamping torque is strong in both monolayer and trilayer thick MoTe$_2$ devices, the observed torque goes to zero in bilayer-thick MoTe$_2$.  

\begin{figure}[t!]
\centering
\includegraphics[width=8 cm]{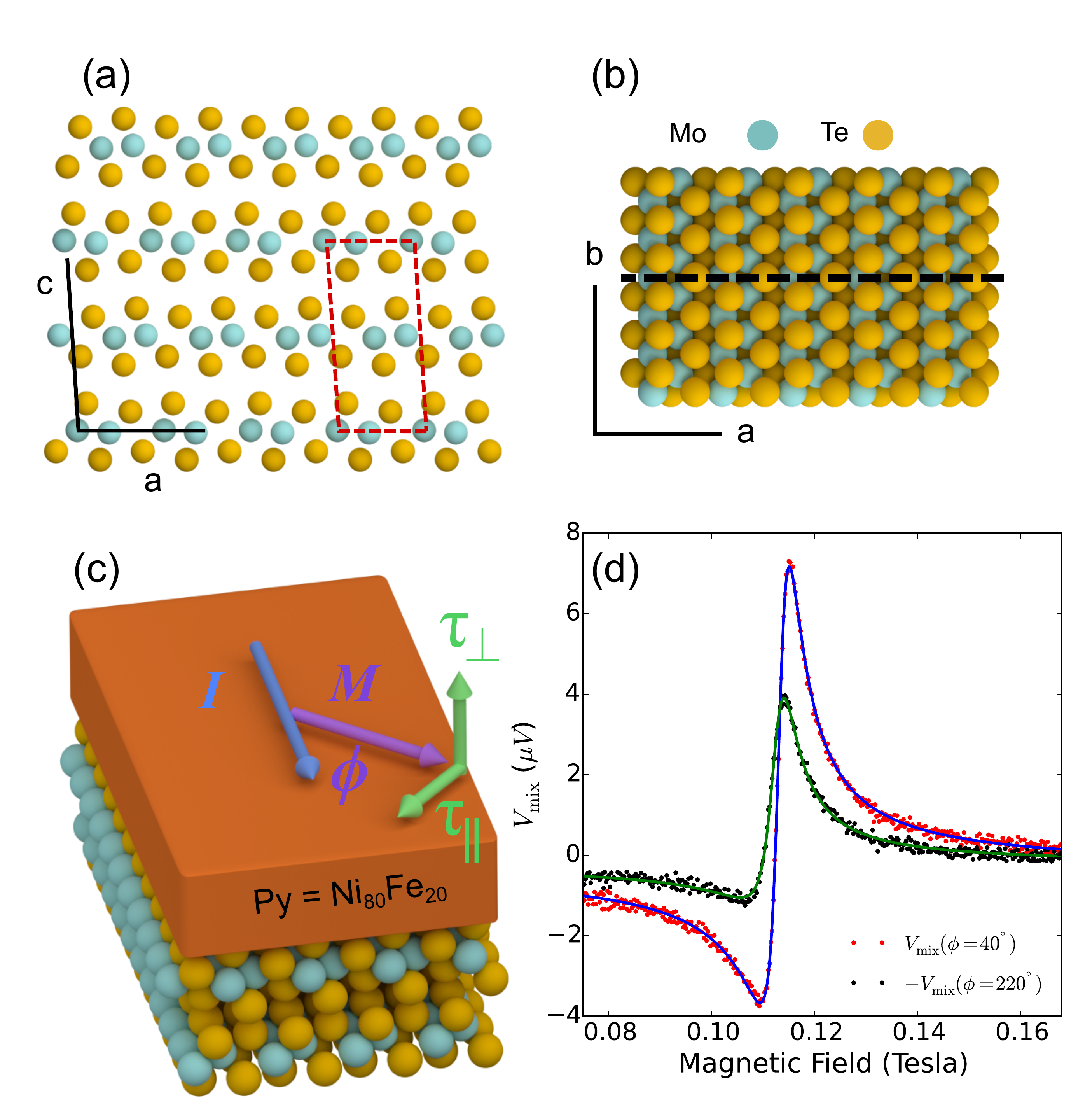}
    \caption{a and b) Structure of the MoTe$_2$ crystal in the monoclinic ($\beta$ or 1$T$') phase (a) depicted in the a-c plane for which the mirror plane is within the page and the Mo chains run into the page and (b) in the a-b plane where the a-c mirror plane is depicted by the dashed black line. c) Geometry of the induced spin-orbit torques in our MoTe$_2$/Py heterostructures. The applied current is defined as being in the $\hat{x}$ direction. d) The measured mixing voltage, V$_{mix}$, as a function of applied magnetic field for Device 1, MoTe$_2$(monolayer) / Py(6 nm), where the current is applied perpendicular to the mirror plane and the field applied at an angle of 40$^{\mathrm{o}}$ (red) and 220$^{\mathrm{o}}$ (black), showing a clear lack of two-fold rotational symmetry in the generated spin-orbit torques. Fits using a sum of symmetric and antisymmetric Lorentzians, shown in blue and green respectively, show good agreement with the data. The applied microwave power is 5 dBm at 9 GHz.}
    \label{MoTe2Fig1}
\end{figure}

The monoclinic ($\beta$ or 1T') phase of MoTe$_2$ provides a unique opportunity to probe the symmetries relevant for the generation of novel spin-orbit torques, in that the individual monolayers of $\beta$-MoTe$_2$ are isostructural to WTe$_2$ monolayers, but are stacked such that inversion symmetry is maintained in the bulk crystal. Bulk $\beta$-MoTe$_2$ has the space group P2$_1$/m, (\#11), with a screw axis along the Mo chain and a mirror plane perpendicular to the screw axis (Fig.\ \ref{MoTe2Fig1}a)\ \cite{Clarke1978PMB}. Similar to WTe$_2$, however, the surface symmetry is limited to just one mirror plane perpendicular to the Mo chain shown in Fig.\ \ref{MoTe2Fig1}b. 

To fabricate our samples we exfoliate flakes of bulk $\beta$-MoTe$_2$ crystal (provided by HQ graphene) onto high resistivity silicon / silicon oxide wafers, where the last step of the exfoliation process is carried out under vacuum ($<10^{-6}$ torr) in the load lock of our sputtering system. We then use grazing angle sputtering to deposit 6 nm of our ferromagnet, Permalloy (Py=Ni$_{80}$Fe$_{20}$), and subsequently cap the films with 2 nm of Al that is oxidized upon exposure to air. The equilibrium direction of the Py magnetic moment is within the sample plane. Flakes are identified for patterning by optical and atomic force microscopy (AFM); we select regions of flakes that are clean (no tape residue) and atomically flat ($<$300 pm roughness) with no monolayer steps over the entire region from which devices will be fabricated. The thicknesses of the $\beta$-MoTe$_2$ flakes can be accurately determined by AFM (with a layer step-height of $\sim$0.7 nm). Samples are then patterned into  device structures using electron beam lithography and ion mill etching. Electrical contact is made using 5 nm Ti/75 nm Pt electrodes.
The data presented in the main text of this work are all taken at room temperature. We have confirmed that all our devices (down to the monolayer limit) are in the $\beta$-MoTe$_2$ phase at room temperature by polarized Raman spectroscopy (see Appendix). We have also used polarized Raman spectroscopy to determine the crystallographic orientation of each device with respect to the applied current direction\ \cite{Beams2016ACSnano,ChenNanoLett2016,ZhouJACS2017,Zhang2016Raman,Wang2017,Cho625,He2018PRB}.

For the ST-FMR measurements\ \cite{Liu2011,Mellnik2014,MacNeill2016}, we use a ground-signal-ground type device structure, in which we apply a GHz frequency current to the MoTe$_2$/Py bar through the capacitive branch of a bias tee. We set the angle of the applied magnetic field with respect to the current direction, $\phi$, and sweep the magnitude of that field to tune the ferromagnet through its resonance condition while measuring the resultant DC mixing voltage at the inductive end of the bias tee. The mixing voltage, $V_{mix}$, as a function of field magnitude can be fit accurately as the sum of symmetric and antisymmetric Lorentzians. The amplitudes of those Lorentzians are related to the in-plane ($\tau_{\parallel}$) and out-of-plane ($\tau_{\perp}$) torques on the ferromagnet, respectively, by\ \cite{Liu2011,Mellnik2014,MacNeill2016}:
\begin{equation}
V_{\mathrm{S}}=-\frac{I_{\mathrm{RF}}}{2}\frac{dR}{d\phi}\frac{1}{\alpha_{\mathrm{G}}\gamma\left(2B_0+\mu_0 M_{\mathrm{eff}}\right)}\tau_{\parallel}\label{eq:VsMoTe2}
\end{equation}
\begin{equation}
V_{\mathrm{A}}=-\frac{I_{\mathrm{RF}}}{2}\frac{dR}{d\phi}\frac{\sqrt{1+\mu_0 M_{\mathrm{eff}}/B_0}}{\alpha_{\mathrm{G}}\gamma\left(2B_0+\mu_0 M_{\mathrm{eff}}\right)}\tau_{\perp},\label{eq:VaMoTe2}
\end{equation}
where $R$ is the device resistance, $dR/d\phi$ is due to the anisotropic magnetoresistance in the Py,\ \ $\mu_0 M_{\mathrm{eff}}$ is the out-of-plane demagnetization field, $B_0$ is the resonance field, $I_{\mathrm{RF}}$ is the microwave current in the bilayer, $\alpha_{\mathrm{G}}$ is the Gilbert damping coefficient and $\gamma$ is the gyromagnetic ratio. Figure \ref{MoTe2Fig1}d shows $V_{\mathrm{mix}}$ at two applied field angles, $40^{\mathrm{o}}$ and $220^{\mathrm{o}}$, for one of our devices (Device 1, containing one monolayer of MoTe$_2$) where the applied current in the device is perpendicular to the MoTe$_2$ mirror plane and the $220^{\mathrm{o}}$ trace has been multiplied by $-1$ for comparison. Fits to the data using a sum of symmetric and antisymmetric Lorentzians show good agreement. 

In high symmetry materials such as Pt, the generated spin-orbit torques are limited by symmetry to consist of an out-of-plane field-like torque, $\vec{\tau_{A}}\propto\hat{m}\times\hat{y}$, and an in-plane antidamping torque, $\vec{\tau_{S}}\propto\hat{m}\times(\hat{m}\times\hat{y})$, which both have a dependence on the magnetization direction $\propto\cos(\phi)$\ \cite{Garello2013}. That the $V_{\mathrm{mix}}$ data in Fig. \ref{MoTe2Fig1}d are not identical up to a minus sign for the two applied field angles indicates that torques in the $\beta$-MoTe$_2$/Py system do not preserve two-fold symmetry, $i.e.$ an out-of-plane antidamping torque, $\vec{\tau_{B}}\propto\hat{m}\times(\hat{m}\times\hat{z})$, may be present. Note that we define the applied current as always being in the $\hat{x}$ direction (Fig.\ \ref{MoTe2Fig1}c). 

\begin{figure}[t!]
\centering
\includegraphics[width=8 cm]{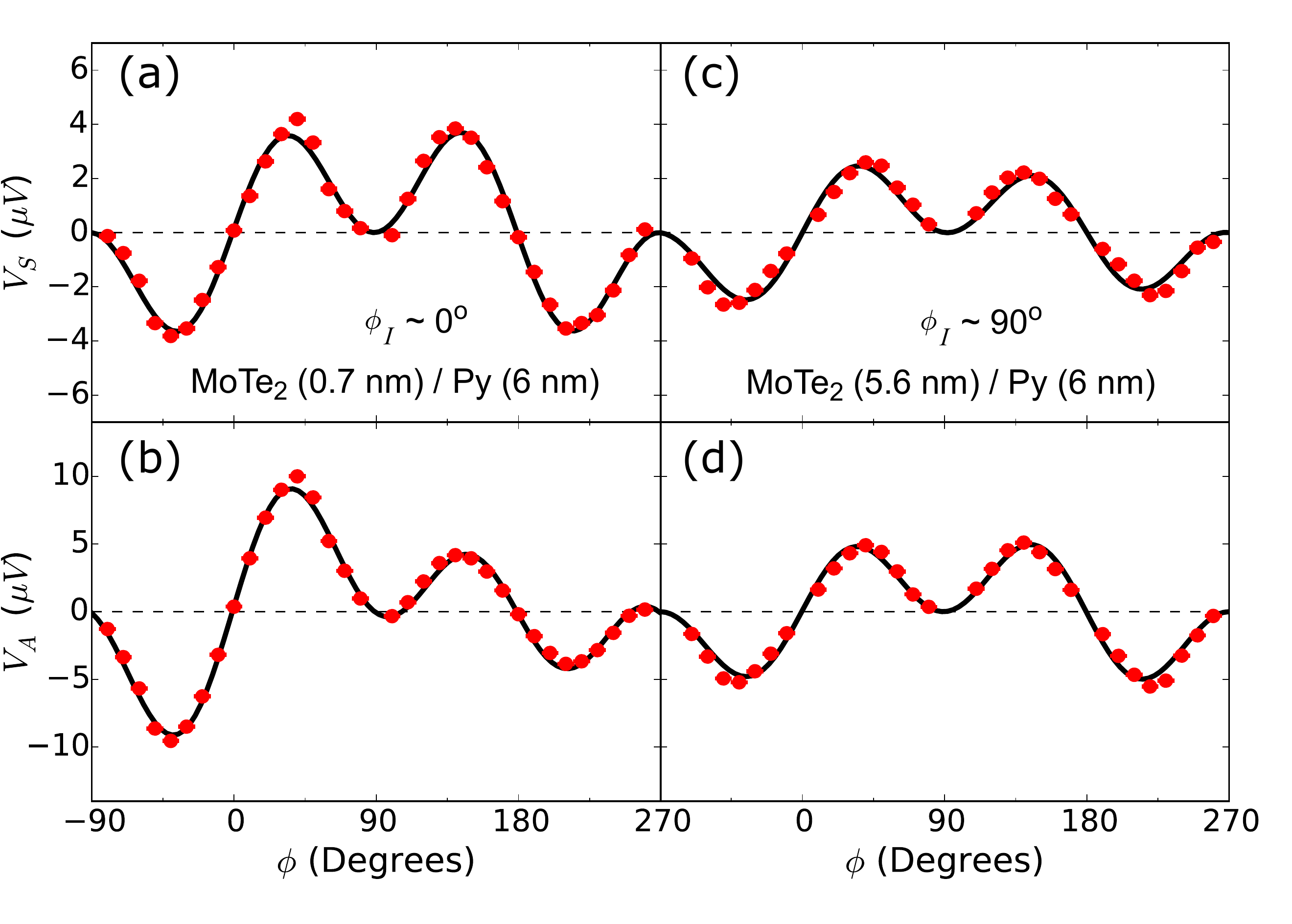}
    \caption{a and b) Dependence on the applied field angle for both the (a) symmetric, $V_{\mathrm{S}}$, and (b) antisymmetric, $V_{\mathrm{A}}$, component of the mixing voltage for Device 1, MoTe$_2$(monolayer) / Py(6 nm) with current applied perpendicular to the MoTe$_2$ mirror plane. Fits of the angular dependence are made using Eqs. \ref{eq:VaAngleMoTe2} and \ref{eq:VsAngleMoTe2}. An out-of-plane antidamping torque is observed, with fit values $B/A=0.302\pm0.001$. c and d) Dependence on the applied field angle for (c) $V_{\mathrm{S}}$ and (d) $V_{\mathrm{A}}$ in Device 2, MoTe$_2$(5.6 nm) / Py(6 nm) with current applied along the MoTe$_2$ mirror plane. No out-of-plane antidamping torque is observed within experimental uncertainty, consistent with the symmetry requirements of the MoTe$_2$ surface. In both samples the applied microwave power is 5 dBm at 9 GHz.}
    \label{MoTe2Fig2}
\end{figure}

To determine the components of current-induced torque, we analyze the extracted fit parameters $V_{\mathrm{S}}$ and $V_{\mathrm{A}}$ as a function of applied field angle.  Representative data for Device 1 are shown in Fig.\ \ref{MoTe2Fig2}a and b.
 If only the torques $\tau_{\mathrm{A}}$ and $\tau_{\mathrm{S}}$ are present, $V_{\mathrm{A}}$ and $V_{\mathrm{S}}$ will be $\propto\sin(2\phi)\cos(\phi)$, where the $\propto\sin(2\phi)$ arises from $dR/d\phi$ due to the anisotropic magnetoresistance of the Py.
However, the angular dependence of $V_{\mathrm{A}}$ (Fig.\ \ref{MoTe2Fig2}b) cannot be described with this simple overall angular dependence $\propto\sin(2\phi)\cos(\phi)$. To extract the other out-of-plane torques present in the system, we fit the angular dependence of $V_{\mathrm{A}}$ as:
\begin{equation}
V_{\mathrm{A}}=\sin(2\phi)[A\cos(\phi)+B+C\sin(\phi)].\label{eq:VaAngleMoTe2}
\end{equation}
The fit parameter $B$ corresponds to torques $\vec{\tau_{B}}\propto\hat{m}\times(\hat{m}\times\hat{z})$. The fit parameter $C$ corresponds to torques $\vec{\tau_{C}}\propto\hat{m}\times\hat{x}$ -- the torque with a Dresselhaus-like symmetry observed in TaTe$_2$ and WTe$_2$ that likely arises from the in-plane resistance anisotropy of the low-symmetry TMD\ \cite{Stiehl2018}. For Device 1, we find a ratio $B/A=0.302\pm0.001$ indicating a sizable out-of-plane antidamping torque, whereas $C$ is zero to within experimental uncertainty. 

We may similarly fit $V_{\mathrm{S}}$ to test for additional in-plane torques:
\begin{equation}
V_{\mathrm{S}}=\sin(2\phi)[S\cos(\phi)+T+U\sin(\phi)],\label{eq:VsAngleMoTe2}
\end{equation}
where $T$ corresponds to torques $\vec{\tau_{T}}\propto\hat{m}\times\hat{z}$, and $U$ gives torques $\vec{\tau_{U}}\propto\hat{m}\times(\hat{m}\times\hat{x})$. In Device 1, $T$ and $U$ are zero within experimental uncertainty. However, other samples show non-zero values for $T$ and $U$ as discussed below.

When a spin-generation system has a single mirror symmetry and the current is applied perpendicular to this mirror plane (as is the case for the MoTe$_2$/Py interface of Device 1 in Fig.\ \ref{MoTe2Fig2}a,b) a net torque generated by an out-of-plane spin is allowed by symmetry (a torque $\propto\hat{m}\times\hat{z}$ or $\propto\hat{m}\times(\hat{m}\times\hat{z})$). However, if the current instead flows along a mirror plane, such a torque is forbidden by symmetry. Figure \ref{MoTe2Fig2}c and d shows $V_{\mathrm{S}}$ and $V_{\mathrm{A}}$ for a MoTe$_2$/Py device in which current is applied along the MoTe$_2$ mirror plane (Device 2). Consistent with this symmetry requirement, fits of $V_{\mathrm{A}}$ using Eq. \ref{eq:VaAngleMoTe2}  yield values of $B$ that are zero within experimental uncertainty. We note the presence of a small, but nonzero, value of $T$ as determined by fits of $V_{\mathrm{S}}$ using Eq. \ref{eq:VsAngleMoTe2}, $T/S=0.067\pm0.003$, which is discussed below and in Appendix\ \ref{mxzappend}.  

The torque conductivity, defined as the angular momentum absorbed by the magnet per second per unit interface area per unit applied electric field, provides an absolute measure of the torques produced in a spin source/ferromagnet heterostructure nominally independent of geometric factors. For a torque $\tau_{\mathrm{K}}$ (where $K$ = $A$, $B$, $C$, $S$, $T$ or $U$) we calculate the corresponding torque conductivity via 
\begin{equation}
\sigma_{\mathrm{K}}=\frac{M_{\mathrm{s}}lwt_{\mathrm{Py}}}{\gamma}\frac{\tau_{\mathrm{K}}}{(lw)E}=\frac{M_{\mathrm{s}}lt_{\mathrm{Py}}}{\gamma}\frac{\tau_{\mathrm{K}}}{I_{\mathrm{RF}}\cdot Z_{\mathrm{RF}}},\label{eq:SigmaK}
\end{equation}
where $M_{\mathrm{s}}$ is the saturation magnetization, $E$ is the electric field, $l$ and $w$ are the length and width of the MoTe$_2$/Py bilayer, $t_{\mathrm{Py}}=6$ nm is the thickness of the Py, and $Z_{\mathrm{RF}}$ is the device impedance. The factor $M_{\mathrm{s}}lwt_{\mathrm{Py}}/\gamma$ is the total angular momentum of the magnet, and converts the normalized torque into units of angular momentum per second. Further details of the ST-FMR analysis can be found in Appendix\ \ref{STFMRappend}. 

\begin{figure}[b!]
\centering
\includegraphics[width=7 cm]{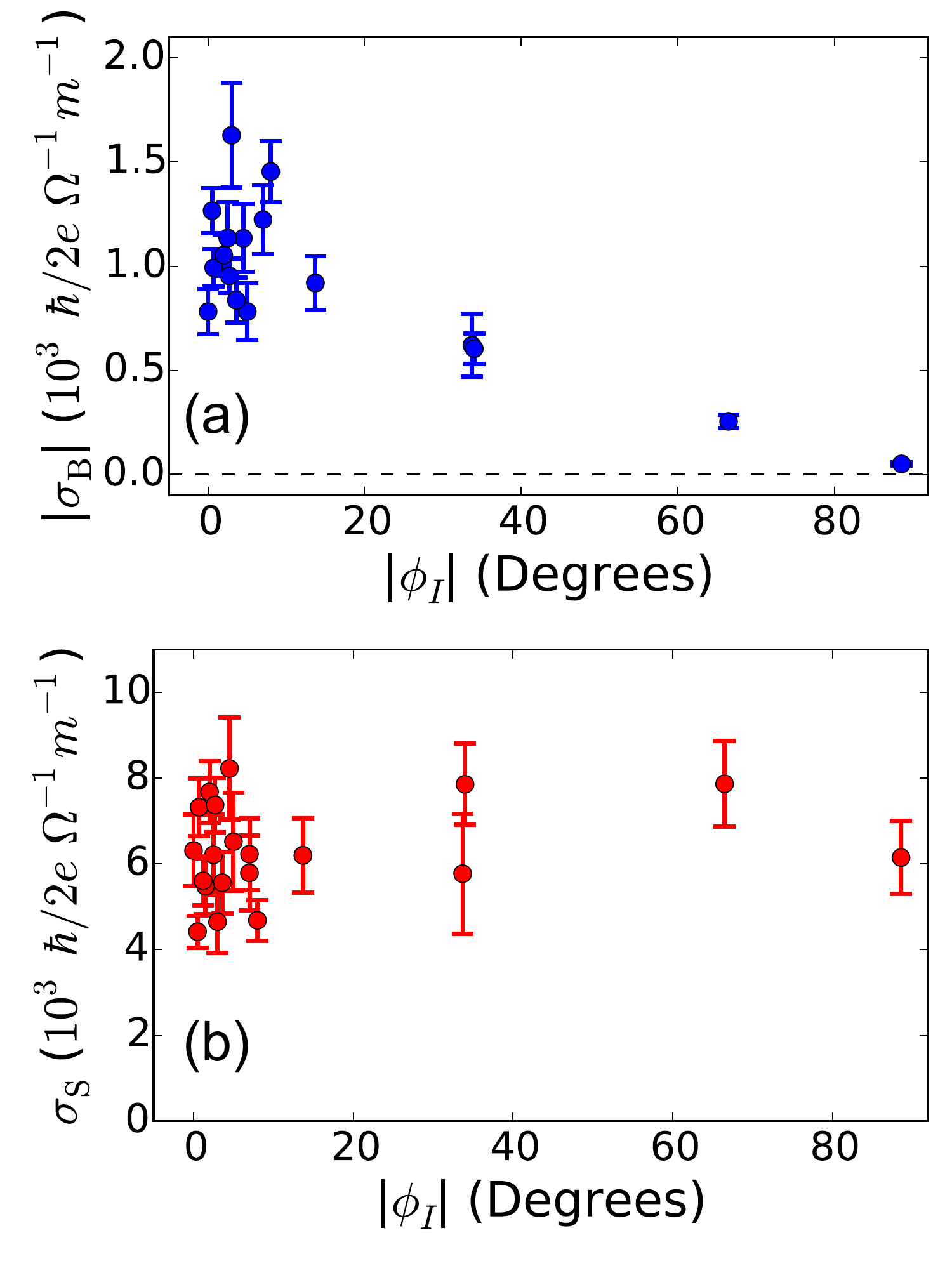}
    \caption{a) Torque conductivities for the out-of-plane antidamping torque, $\tau_{\mathrm{B}}$, as a function of the angle $|\phi_I|$ for 17 of our MoTe$_2$/Py devices, all with distinct MoTe$_2$ thicknesses. We have excluded our bilayer MoTe$_2$ devices in this plot, which are discussed in detail later. b) Torque conductivities for the standard in-plane antidamping torque, $\tau_{\mathrm{S}}$, as a function of $|\phi_I|$ in all of our MoTe$_2$/Py devices. In both plots the applied microwave power is 5 dBm. Torque conductivities are averaged over measurements at frequencies 8-11 GHz in steps of 1 GHz.}
    \label{MoTe2Fig3}
\end{figure}

We have determined the torque conductivities for 20 MoTe$_2$($t_{\mathrm{TMD}}$)/Py(6 nm) devices, all with distinct thicknesses of MoTe$_2$, $t_{\mathrm{TMD}}$, and angles between the current direction and the MoTe$_2$ mirror plane. Details of each device can be found in Table\ \ref{tab:MoTe2} of the Appendix.
We define $\phi_{I}$ as the angle between the current and the vector normal to the MoTe$_2$ mirror plane (typically called the b-axis in the $\beta$ phase), such that $\phi_{I}=0^{\mathrm{o}}$ is perpendicular to the mirror plane and $\phi_{I}=90^{\mathrm{o}}$ is parallel. Figure \ref{MoTe2Fig3}a shows $\sigma_{\mathrm{B}}$ as a function of $\phi_{I}$ for 17 of our devices (we have excluded our bilayer thick MoTe$_2$ devices for now, which will be discussed later). Consistent with the symmetry requirements on the torques, $\sigma_{\mathrm{B}}$ is largest when current is applied perpendicular to the MoTe$_2$ mirror plane and is progressively reduced as more of the applied current flows along the mirror plane. Note that in Fig.\ \ref{MoTe2Fig3}a we have plotted $|\sigma_{\mathrm{B}}|$. This is because the sign of $\sigma_{\mathrm{B}}$ is not solely determined by $\phi_{I}$ but also the canting of the molybdenum dimerization at the MoTe$_2$/Py interface, which we do not control and cannot determine by polarized Raman spectroscopy (to visualize this difference, consider a two-fold rotation about the MoTe$_2$ c-axis for a monolayer).

In contrast to the strong dependence on $\phi_{I}$ for $\sigma_{\mathrm{B}}$, the torque conductivity for the conventional component of in-plane antidamping spin Hall torque, $\sigma_{\mathrm{S}}$, shows no significant dependence (Fig.\ \ref{MoTe2Fig3}b). This is similar to the $\sigma_{\mathrm{S}}$ dependence on $\phi_{I}$ observed in our WTe$_2$/Py heterostructures\cite{MacNeill2016,MacNeill2017}. We note, however, that the relative insensitivity of $\sigma_{\mathrm{S}}$ to the in-plane current direction observed in $\beta$-MoTe$_2$ and WTe$_2$ is not required by symmetry, and in general the magnitudes of the in-plane spins generated in response to a current along the a or b-axes are allowed to differ\cite{Zhu2018PRB}. We obtain an average value of  $\sigma_{\mathrm{S}}$ for our MoTe$_2$/Py devices of $5800\pm160$ $\hbar/(2e)$ $(\Omega^{-1}m^{-1})$, smaller than the average value observed in our WTe$_2$/Py heterostructures, $8000\pm200$ $\hbar/(2e)$ $(\Omega^{-1}m^{-1})$\ \cite{MacNeill2016,MacNeill2017}, and larger than the $\approx3000$ $\hbar/(2e)$ $(\Omega^{-1}m^{-1})$ observed in our NbSe$_2$/Py heterostructures\ \cite{Guimaraes2018Nano}. 

\begin{figure}[t!]
\centering
\includegraphics[width=7 cm]{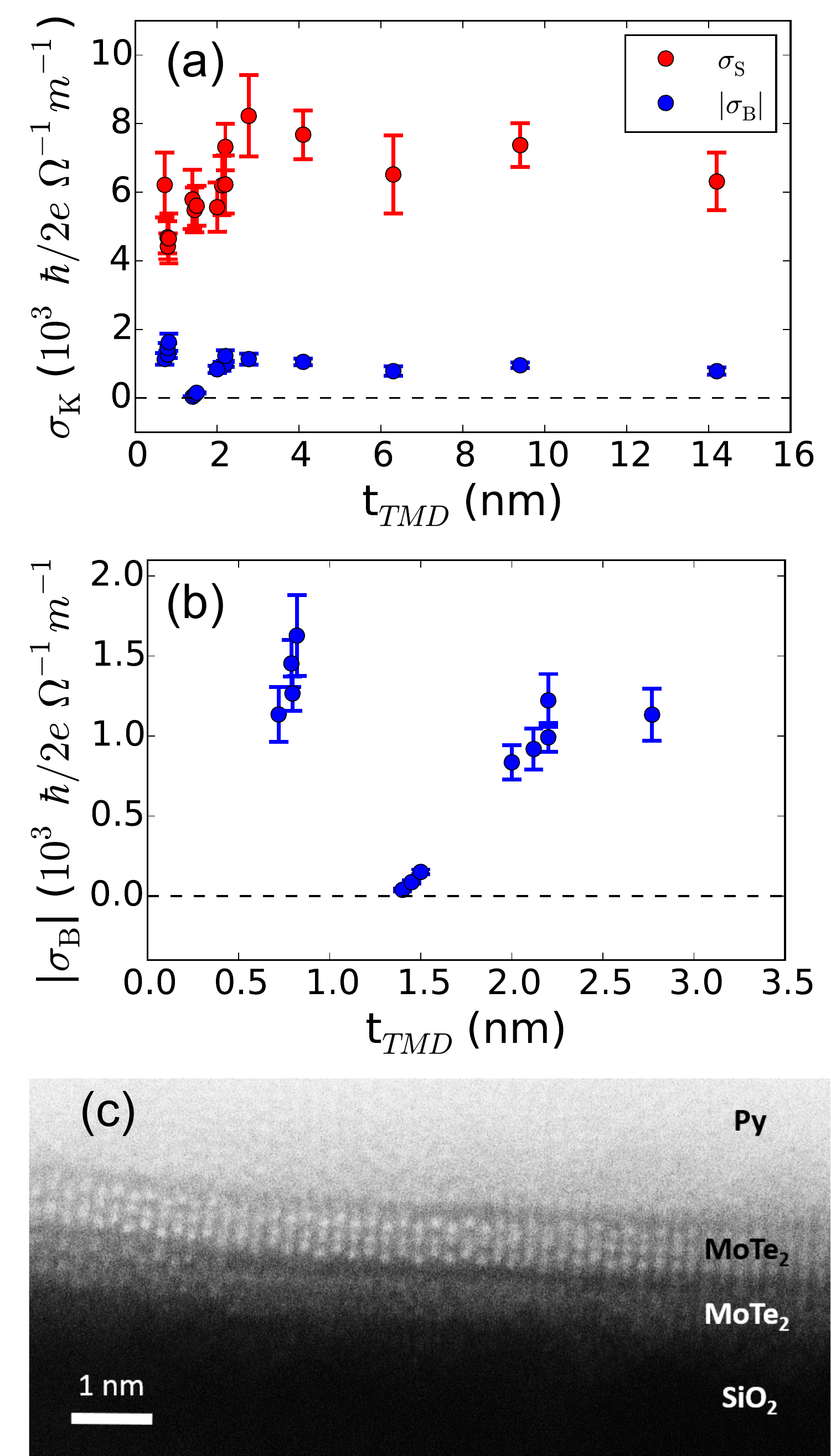}
    \caption{a) The torque conductivities $\sigma_{\mathrm{B}}$ and $\sigma_{\mathrm{S}}$ as a function of TMD thickness for devices with current aligned perpendicular to the MoTe$_2$ mirror plane. b) Detail of the thickness dependence of $\sigma_{\mathrm{B}}$ in the monolayer to quadlayer regime (with a layer step size of $\sim$0.7 nm). The devices that show a value near zero are bilayer MoTe$_2$. The applied microwave power is 5 dBm. Torque conductivities are averaged over measurements at frequencies 8-11 GHz in steps of 1 GHz. c) Cross-sectional HAADF-STEM image of a MoTe$_2$(1.4 nm)/Py device looking down the $\beta$-MoTe$_2$ a-axis and along the b-axis. The image shows two MoTe$_2$ layers, where the layer interfaced with the Si/SiO$_2$ substrate (white MoTe$_2$ label) shows regions of disorder which we attribute to partial oxidation. The Py is polycrystalline.}
    \label{MoTe2Fig5}
\end{figure}

To help analyze the mechanism that drives the spin-orbit torques in our MoTe$_2$/Py heterostructures, it is helpful to study the torques as a function of MoTe$_2$ thickness, holding the crystal alignment fixed. In Appendix\ \ref{OerstedAppend}, we discuss the observed thickness dependence for $\sigma_{\mathrm{A}}$, and show that this torque contribution is dominated by the Oersted field.
Figure \ref{MoTe2Fig5}a shows the thickness dependence for $\sigma_{\mathrm{B}}$ and $\sigma_{\mathrm{S}}$ for devices with current aligned perpendicular to the MoTe$_2$ mirror plane, where $|\phi_{I}|<15^{\circ}$ and usually less than 10$^{\circ}$. 
Both torques are largely independent of MoTe$_2$ thickness, with the notable exception of bilayer MoTe$_2$ devices, implying an interfacial origin for these torque components. This is in qualitative similarity to the TMD thickness dependence previously observed in WTe$_2$. The striking exception in the thickness dependence is from our devices in which the MoTe$_2$ is just a bilayer (1.4 nm) thick (3 different samples). In these devices, no out-of-plane antidamping torque is observed within our experimental uncertainty. Excluding the bilayer devices, we find an average value for $|\sigma_{\mathrm{B}}|=1020\pm30$ $\hbar/(2e)$ $(\Omega^{-1}m^{-1})$. It is interesting that while the magnitude of $\sigma_{\mathrm{S}}$ in MoTe$_2$ is similar to that observed in our WTe$_2$ devices, the value of $|\sigma_{\mathrm{B}}|$ is approximately 1/3 that of WTe$_2$\ \cite{MacNeill2016,MacNeill2017}.

Figure \ref{MoTe2Fig5}b shows in more detail the measured out-of-plane antidamping torque conductivity $|\sigma_{\mathrm{B}}|$ in MoTe$_2$/Py heterostructures in monolayer steps from a single MoTe$_2$ layer to quadlayer MoTe$_2$, all with current perpendicular to the MoTe$_2$ mirror plane.  The contrast is striking between the large torques in the monolayer, trilayer and quadlayer MoTe$_2$ devices, and the nearly-zero torque in the three bilayer samples. In our previous work on WTe$_2$/Py devices, bilayer WTe$_2$ (two samples) also showed a decrease in $\sigma_{\mathrm{B}}$, although in that case $\sigma_{\mathrm{B}}$ for the bilayer devices was $\mathtt{\sim}1/2$ that of the monolayer and trilayer devices rather than zero (see Fig.\ 5 of Ref.\ \cite{MacNeill2017}). The origin of these reductions is unknown. Only the out-of-plane antidamping torque is reduced, not $\sigma_{\mathrm{S}}$ or $\sigma_{\mathrm{A}}$. In WTe$_2$, non-symmorphic crystal symmetries (the b-c glide plane and c-axis screw) require that the spin responsible for generating $\tau_{\mathrm{B}}$ must have opposite signs in adjacent layers\ \cite{MacNeill2017}. While $\beta$-MoTe$_2$ does not possess the same non-symmorphic symmetries, there is an effective in-plane  polar vector at the $\beta$-MoTe$_2$/Py interface that changes sign for adjacent $\beta$-MoTe$_2$ layers\ \cite{Shi2019}, which could lead to oppositely-directed current-induced out-of-plane spins in adjacent layers. Such a layer-dependent sign for the out-of-plane spin might lead to a partial cancellation of contributions from adjacent layers, and may have some bearing on this layer-dependent effect. However, this mechanism alone is difficult to reconcile with our observations that a strong layer dependence is present only in the bilayer MoTe$_2$, and that the lower layer in our bilayer devices is also likely partially oxidized (see below).

The in-plane field-like torque, $\vec{\tau}_{\mathrm{T}}\propto\hat{m}\times\hat{z}$, has the same symmetry constraints as $\vec{\tau}_{\mathrm{B}}$ and is symmetry allowed in our MoTe$_2$/Py devices when the current is applied perpendicular to the MoTe$_2$ mirror plane. Interestingly, we observe significant values of $\sigma_{\mathrm{T}}$ only in devices with sufficiently thick MoTe$_2$, above about 3 nm. Details can be found in Appendix\ \ref{mxzappend} and\ \ref{tempSOTappend}. In all such devices the ratio of $\sigma_{\mathrm{T}}/\sigma_{\mathrm{B}}$ is always negative, even though the signs of $\sigma_{\mathrm{T}}$ and $\sigma_{\mathrm{B}}$ vary from sample to sample. In addition, we find that $\tau_{\mathrm{T}}$ and $\tau_{\mathrm{B}}$ scale similarly with the sample temperature. These observations suggest that the two torque components are correlated. However, we also find that $\sigma_{\mathrm{T}}$ and $\sigma_{\mathrm{B}}$ have distinct dependencies on the MoTe$_2$ thickness, in tension with the idea that the two torques are generated through the same mechanism. Surprisingly, we also find a significant, though reduced, value of $\sigma_{\mathrm{T}}$ in samples where current is flowed predominately along the MoTe$_2$ mirror plane (Fig.\ \ref{MoTe2Fig2}c) -- $i.e$ the condition under which the torque is symmetry forbidden. This suggests the possibility of two distinct mechanisms contributing to the generation of $\tau_{\mathrm{T}}$: one which is correlated with $\tau_{\mathrm{B}}$ and the MoTe$_2$ crystal structure, and the other independent of the nominal MoTe$_2$ crystal structure and possibly due to strain induced by the fabrication procedure\ \cite{Guimaraes2018Nano,Lee2017}. We note that the observation of non-zero values of both $\sigma_{\mathrm{B}}$ and $\sigma_{\mathrm{T}}$ in the same sample is in contrast to previous measurements on other TMDs. In WTe$_2$ we did not observe any significant value of $\tau_{\mathrm{T}}$ even though a large $\tau_{\mathrm{B}}$ was present (see Fig.\ 6 of Ref.\ \cite{MacNeill2017}). For (presumably) strained NbSe$_2$/Py devices the situation was exactly the opposite. There we observed a large value of $\tau_{\mathrm{T}}$, but no $\tau_{\mathrm{B}}$, though with no clear dependence on the TMD thickness. 

To better understand the microscopic structure of the MoTe$_2$/Py samples, we have performed cross-sectional high-angle annular dark-field scanning transmission electron microscopy (HAADF-STEM) imaging on a bilayer $\beta$-MoTe$_2$/Py device (Fig.\ \ref{MoTe2Fig5}c) as initially identified by AFM. Both $\beta$-MoTe$_2$ layers are clearly visible. However, the layer adjacent to the Si/SiO$_2$ substrate shows increased localized disorder in some regions, which we attribute to partial oxidation. Note that this partially oxidized layer remains crystallographically oriented in the unoxidized regions, while the top layer appears pristine. Oxidation just of the lower MoTe$_2$ layer is consistent with our sample fabrication process -- the layer of MoTe$_2$ adjacent to the substrate is exposed to air before being placed on the substrate for exfoliation, while the other layers are protected from air exposure throughout the fabrication process. The existence of partial oxidation of the lowest MoTe$_2$ layer is also consistent with the signal strengths from our polarized Raman spectroscopy measurements for monolayer and bilayer thick $\beta$-MoTe$_2$/Py devices: the bilayer thick samples show a strong signal, whereas the monolayer devices show a significantly weaker signal, but with no evidence of crystallographic twinning in either data set (see Appendix).

In summary, we have studied the current-induced spin-orbit torques in $\beta$-MoTe$_2$/Py heterostructures at room temperature. We have observed an out-of-plane antidamping torque, $\tau_{\mathrm{B}}$, qualitatively similar to the $\tau_{\mathrm{B}}$ observed in WTe$_2$/Py heterostructures. This torque is consistent with the symmetries of the MoTe$_2$ surface -- at the interface of MoTe$_2$ and Py the structural symmetries are limited to a single mirror plane, and consistent with that symmetry, $\tau_{\mathrm{B}}$ is observed only when a component of the current is applied perpendicular to that mirror plane. This demonstrates that breaking of inversion symmetry in the bulk of the spin-generation layer is not a necessary requirement for $\tau_{\mathrm{B}}$. The magnitude of the observed torque conductivity is $\mathtt{\sim}$1/3 that observed in similar devices that use WTe$_2$ as the spin source layer. In both materials $\sigma_{\mathrm{B}}$ is largely independent of the TMD thickness. The torque conductivity for the standard antidamping torque, $\sigma_{\mathrm{S}}$, is also independent of thickness indicating that both torques are likely generated by an interfacial mechanism. The notable exception in the thickness dependence of $\sigma_{\mathrm{B}}$ is for bilayer-thick MoTe$_2$ devices, for which $\sigma_{\mathrm{B}}$ is zero to within measurement uncertainty. Bilayer WTe$_2$ devices also have a greatly-reduced out-of-plane antidamping torque compared to monolayer or trilayer devices, but the origin of this effect is unknown. 

Acknowledgements: GMS acknowledges useful conversations with Dr. David MacNeill. The primary support for this project came from the US Department of Energy (DE-SC0017671). Electron microscopy was performed with support from the NSF through PARADIM as part of the Materials for Innovation Platform Program. Sample fabrication was performed in the shared facilities of the Cornell Center for Materials Research (NSF DMR-1719875) and at the Cornell Nanoscale Science \& Technology Facility, part of the National Nanotechnology Coordinated Infrastructure, which is supported by the NSF (NNCI-1542081). The FEI Titan Themis 300 was acquired through Grant NSF-MRI-1429155, with additional support from Cornell University, the Weill Institute, and the Kavli Institute at Cornell.

\appendix

\begin{table*}[!tbhp]
\centering
\scalebox{0.85}{
    \begin{tabular}{|c| c| c| c| c| c| c|}
    \hline
      Device  & $t$ (nm)  & $L\times W$ ($\mu$m) & $\sigma_{\mathrm{S}}$ &  $\sigma_{\mathrm{B}}$ & $\sigma_{\mathrm{T}}$ & $|\phi_{\mathrm{I}}|$ (degrees)  \\ 
      Number& $\pm$ 0.3 nm & $\pm$ 0.2 $\mu$m &$(10^{3}$ $\hbar/2e$ $\Omega^{-1}m^{-1})$ &$(10^{3}$ $\hbar/2e$ $\Omega^{-1}m^{-1})$ &$(10^{3}$ $\hbar/2e$ $\Omega^{-1}m^{-1})$ & $\pm 5^{\circ}$  \\ \hline
   1 & 0.7 & $5\times4$    & -6.2(1)  & -1.1(2)  & 0.026(5) & 3  \\ \hline
   2 & 5.6 & $5\times4$     & -6.1(9)  & 0.051(7) & -0.39(6) & 89  \\  \hline
   3 & 6.3 & $5\times4$     &  -7(1)   & -0.8(1)  & 0.8(1)   & 5 \\ \hline
   4 & 1.4 & $5\times4$     &  -5.8(9) & -0.04(1) & -0.22(5) & 7 \\ \hline
   5 & 14.2 & $5\times4$    &  -6.3(8) & -0.8(1)  & 0.7(1)   & 0 \\ \hline
   6 & 2.1 & $5\times4$     &  -6.2(9) & 0.9(1)   & -0.13(2) & 14 \\ \hline
   7 & 2.2 & $5\times4$     &  -7.3(7) & -0.99(9) & 0.032(5) & 1 \\ \hline
   8 & 2.8 & $5\times4$     &  -8(1)   & -1.1(2)  & 0.28(4)  & 5 \\ \hline
   9 & 4.1 & $5\times4$     &  -7.7(7) & 1.1(1)   & -0.78(7) & 2 \\ \hline
   10 & 9.4 & $5\times4$   & -7.4(6)  & 0.95(8)  & -1.0(1)  & 3 \\ \hline
   11 & 0.8 & $5\times4$    & -4.4(4)  & -1.3(1)  & 0.14(1)  & 1 \\ \hline
   12 & 1.5 & $5\times4$    & -5.5(7)  &-0.09(1)  & 0.25(3)  & 2 \\ \hline
   13 & 1.5 & $5\times4$    & -5.6(6)  &-0.15(2)  & 0.23(2)  & 1 \\ \hline
   14 & 2.0 & $5\times4$    & -5.6(7)  &-0.8(1)   & 0.21(3)  & 4 \\ \hline
   15 & 2.2 & $5\times4$   & -6.2(8)  & 1.2(2)   & 0.18(2)  & 7 \\ \hline
   16 & 0.8 & $5\times4$   & -4.7(5)  &-1.4(1)   & 0.06(1)  & 8 \\ \hline
   17 & 0.8 & $5\times4$   & -4.7(7)  &1.6(3)    & 0.08(1)  & 3 \\ \hline
   18 & 2.2 & $5\times4$    & -5(1)    & 0.6(2)   & -0.17(4) & 34 \\ \hline
   19 & 2.3 & $4.5\times4$  & -8(1)    & -0.60(7) & -0.41(5) & 34 \\ \hline
   20 & 9.4 & $4\times3$    & -8(1)    & -0.26(3) & -0.55(7) & 67 \\ \hline

    \end{tabular}}
    \caption{Comparison of device parameters, and torque conductivities for MoTe$_2$/Py bilayers. Here $\phi_{I}$ is the angle between the current and the crystal axis perpendicular to the MoTe$_2$ mirror plane as measured by polarized Raman spectroscopy. For the torque conductivities, the () gives the uncertainty of the last reported digit. }\label{tab:MoTe2}
\end{table*}

\section{Calibrated ST-FMR measurements}\label{STFMRappend}

To make a quantitative measurement of the magnitude of the torques using Eqs. \ref{eq:VsMoTe2} and \ref{eq:VaMoTe2} we must first determine values for $\alpha_{\mathrm{G}}$, $R(\phi)$, and $I_{\mathrm{RF}}$. The Gilbert damping is estimated from the frequency dependence of the linewidth via $\Delta=2\pi f\alpha_{\mathrm{G}}/\gamma+\Delta_0$, where $\Delta_0$ is the inhomogeneous broadening. $R(\phi)$ is determined by measurements of the device resistance as a function applied in-plane magnetic field angle (with a field magnitude of 0.1 T). The RF current is determined by calibrating the reflection coefficients of our devices ($S_{11}$) and the transmission coefficient of our RF circuit ($S_{21}$) through vector network analyzer measurements. These calibrations allow calculation of the RF current flowing in the device as a function of applied microwave power and frequency:
\begin{equation}
I_{\mathrm{RF}}=2\sqrt{1 \mathrm{mW}\cdot 10^{\frac{P_{\mathrm{s}}(\mathrm{dBm})+S_{21}(\mathrm{dBm})}{10}}(1-|\Gamma|)^2/50 \Omega}
\end{equation}
where $P_{\mathrm{s}}$ is the power sourced by the microwave generator and $\Gamma=10^{S_{11}(\mathrm{dBm})/20}$. The frequency dependent device impedance, $Z_{\mathrm{RF}}$, is given by $50$ $\Omega$ $\cdot$ $(1+\Gamma)/(1-\Gamma)$. 

In order to determine a torque conductivity (Eq. \ref{eq:SigmaK}), we must also obtain a value of $M_{\mathrm{s}}$ for the Py. As $M_{\mathrm{s}}$ is influenced by the material on which the Py grows (here, MoTe$_2$), and as mm-scale $\beta$-MoTe$_2$/Permalloy heterostructures are unavailable, we are unable to measure $M_{\mathrm{s}}$ directly via magnetometry. Instead we approximate $M_{\mathrm{s}}\approx M_{\mathrm{eff}}$, which we have found to be accurate in other Py heterostructure systems. We estimate an average value of $\mu_{\mathrm{o}}M_{\mathrm{eff}}=0.84\pm0.01$ T as extracted from the ST-FMR measurements.

\section{Devices Parameters}
Table \ref{tab:MoTe2} shows a comparison of device parameters and torque conductivities for all samples presented in this work.

\section{Determination of Crystal Orientation}\label{MoTe2CrystAxisDetermine}
Crystals of $\beta$-MoTe$_2$ exfoliate in the a-b plane and are generally elongated in the Mo-chain direction, with sharp and cleanly cleaved edges running parallel to that direction. This is very similar to WTe$_2$, and can be used as a first-order approximation of the in-plane crystal axis during device fabrication.  We have also verified the crystal orientation more precisely for each of our devices using Raman spectroscopy.  

\begin{figure*}[t!]
\centering
\includegraphics[width=12 cm]{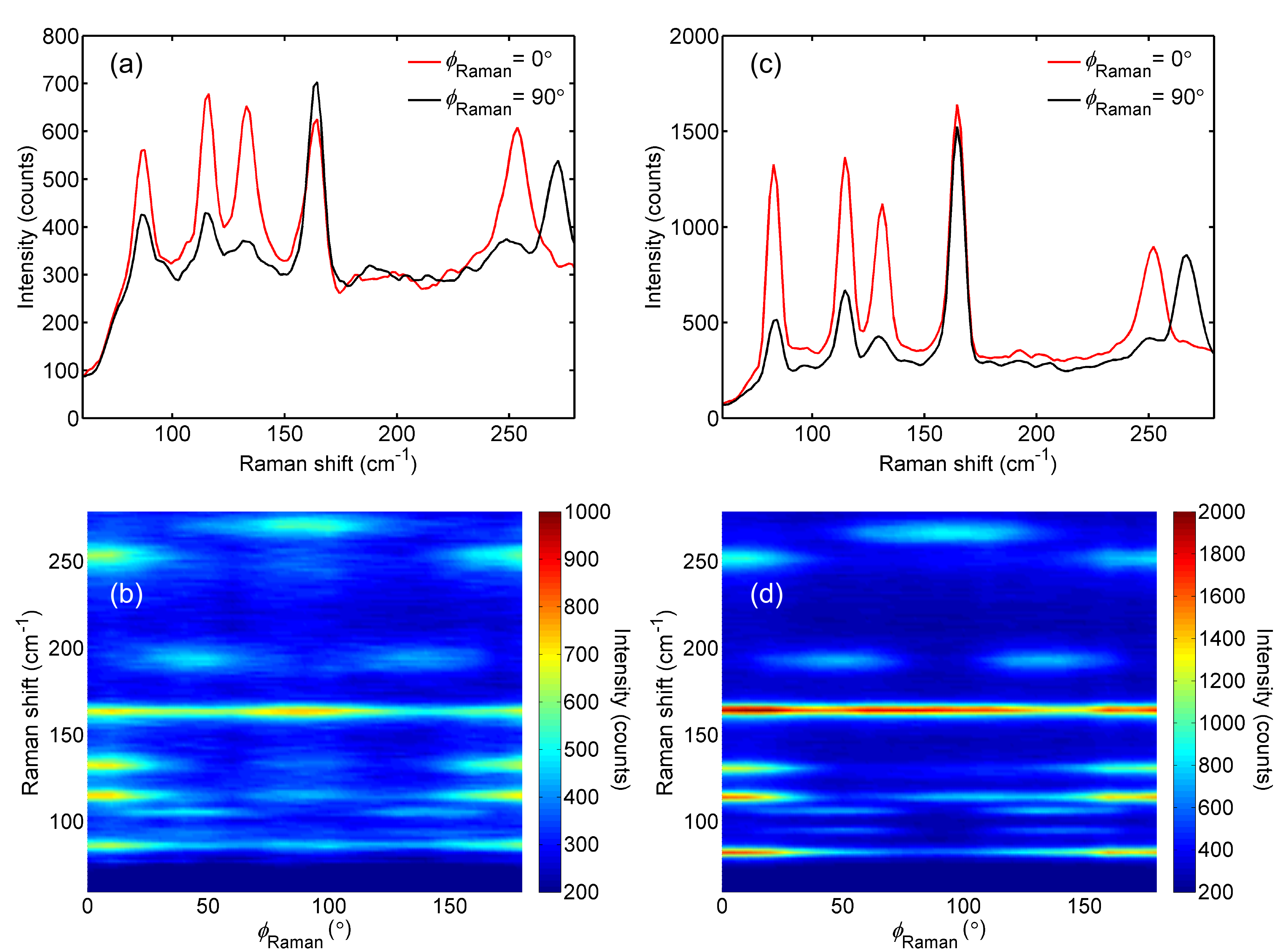}
    \caption{a,c) Raman spectra for MoTe$_2$ monolayer and bilayer / Py devices respectively, with a linearly polarized 488 nm excitation aligned parallel to a polarizer in front of the spectrometer. ${\phi _{{\rm{Raman}}}}$ is the angle between the excitation polarization and the device current direction (along the bar).  The red traces show spectra with the polarization parallel to the current and the black traces show spectra with the polarization perpendicular. b,d) Angular dependence of the Raman spectra for the two devices. The color map represents the peak intensity (with units of counts), where both spectra are taken under the same excitation power. The maximum of the $ \sim 133$ cm$^{-1}$ and $ \sim 250$ cm$^{-1}$ peaks correspond to the MoTe$_2$ b-axis, where ${\phi _{{\rm{Raman}}}} \to  - {\phi _{I}}$.}
    \label{MoTe2PolRaman}
\end{figure*}

\subsection{Raman Spectroscopy}
The Raman spectra of MoTe$_2$ has been well characterized in all three crystal phases\ \cite{Beams2016ACSnano,ChenNanoLett2016,ZhouJACS2017,Zhang2016Raman,Wang2017,Cho625,He2018PRB}. We performed Raman measurements with a confocal Raman microscope using a linearly polarized 488 nm excitation and a parallel polarizer placed in front of the spectrometer. The sample is aligned to the linear polarization direction (along the length of the device) and spectra are taken as the sample is rotated in steps of 10$^{\mathrm{o}}$. The maximum of the $ \sim 133$ cm$^{-1}$ and $ \sim 250$ cm$^{-1}$ peaks correspond to the MoTe$_2$ b-axis. Figure\ \ref{MoTe2PolRaman} shows polarized Raman spectra for two MoTe$_2$/Py devices in which the MoTe$_2$ is (a,b) a monolayer and (c,d) a bilayer thick. The symmetries of the observed peaks are consistent with previous measurements\ \cite{Beams2016ACSnano,ChenNanoLett2016}. No evidence of crystallographic twinning is observed in the polarized Raman spectra. 

\subsection{Magnetic Easy Axis}
In WTe$_2$/Py bilayers, we observed previously that the WTe$_2$ induced a strong in-plane magnetic easy axis that corresponded with the b-axis of the crystal, regardless of the applied current direction\ \cite{MacNeill2016,MacNeill2017}. This correlation provided an efficient means for extracting the angle between the WTe$_2$ crystal axes and the current direction through  electrical measurements alone.  However, we find that MoTe$_2$ does not induce any significant magnetic easy-axis within the Py, so this method cannot be used to characterize the crystal alignment of MoTe$_2$.  TaTe$_2$ generates a magnetic easy axis with strength intermediate between WTe$_2$ and MoTe$_2$\ \cite{Stiehl2018}. This variation suggests different degrees of coupling between the TMDs and Permalloy. 

\section{Out-Of-Plane Field-Like Torque and Oersted Torque}\label{OerstedAppend}
\begin{figure}[b!]
\centering
\includegraphics[width=8 cm]{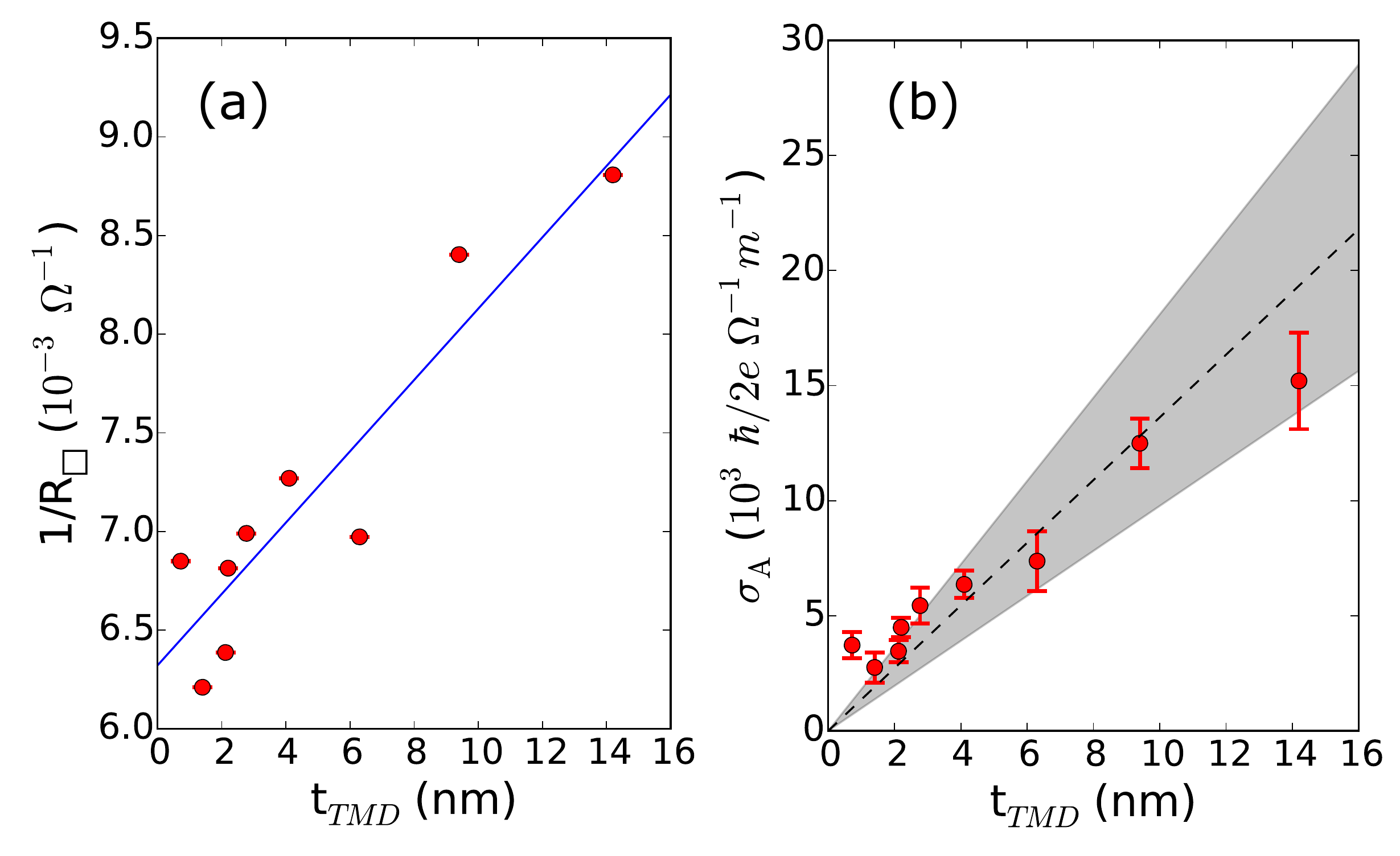}
    \caption{a) Inverse sheet resistance of devices with current perpendicular to the MoTe$_2$ mirror plane (red circles) as a function of MoTe$_2$ thickness as measured by a two-point method. The blue line gives a fit using Eq. \ref{eq:RsquareVt} to extract the sheet resistances for the Py and MoTe$_2$. b) Torque conductivity for the out-of-plane field like torque ($\propto \hat{m}\times\hat{y}$) as a function of thickness for devices with current perpendicular to the MoTe$_2$ mirror plane (red circles). The dashed line gives the predicted Oersted field contribution from the torques (Eq. \ref{eq:SigmaOe}) and the shaded region gives the spread in the expected contribution as given by the uncertainty in the measured charge conductivity of MoTe$_2$. The applied microwave power is 5 dBm. Torque conductivities are averaged over measurements at frequencies 8-11 GHz in steps of 1 GHz. In both plots, we have used only devices in which the Py thin-film and Al capping layers were grown in the same sputtering deposition batch in order to minimize effects from small variations in the Py and Al thicknesses.}
    \label{MoTe2Fig4}
\end{figure}
We extract the individual resistivities of the MoTe$_2$ and Py layers using the two-point resistances of our devices for which the current is aligned perpendicular to the MoTe$_2$ mirror plane ({\it i.e.}, $\phi_I = 0^{\mathrm{o}}$) within 15$^{\circ}$ and usually within 10$^{\circ}$, where we have used only devices in which the Py thin-film and Al capping layers where grown in the same sputtering deposition batch in order to minimize effects from small variations in the Py and Al thicknesses. We plot the inverse of the sheet resistance as a function of $t_{\mathrm{TMD}}$ (Fig.\ \ref{MoTe2Fig4}a), and using the relation:
\begin{equation}
\frac{1}{R_{\square}}=\frac{l}{wR}=\frac{t_{\mathrm{Py}}}{\rho_{\mathrm{Py}}}+\frac{t_{\mathrm{TMD}}}{\rho_{\mathrm{TMD}}},\label{eq:RsquareVt}
\end{equation}
extract the resistivities $\rho_{\mathrm{Py}}=95\pm2$ ($\mu\Omega$ $cm$) and $\rho_{\mathrm{TMD}}=550\pm75$ ($\mu\Omega$ $cm$). The value obtained for $\rho_{\mathrm{Py}}$ is similar to that seen in our WTe$_2$/Py devices when the Py is deposited using glancing-angle sputtering.

Figure \ref{MoTe2Fig4}b shows $\sigma_{\mathrm{A}}$ as a function of $t_{\mathrm{TMD}}$ (red circles) for devices in which current is aligned perpendicular to the MoTe$_2$ mirror plane. A strong thickness dependence of the torques is observed. In many material systems this component of torque, $\propto\hat{m}\times\hat{y}$ (for current in the $\hat{x}$ direction), is dominated by the Oersted torque -- that is, the magnetic field generated from a simple current-carrying wire. For instance, we have previously shown that this is the case in the WTe$_2$/Py and NbSe$_2$/Py systems. We can model the torque conductivity generated by the Oersted torque as:
\begin{equation}
\sigma_{\mathrm{Oe}}=\frac{e \mu_{\mathrm{o}}M_{\mathrm{S}}t_{\mathrm{Py}}\sigma_{\mathrm{TMD}}}{\hbar}t_{\mathrm{TMD}},\label{eq:SigmaOe}
\end{equation} 
where $\sigma_{\mathrm{TMD}}$ is the charge conductivity of the MoTe$_2$. The dashed line in Fig.\ \ref{MoTe2Fig4}b shows the predicted Oersted torque using the extracted value of $\rho_{\mathrm{TMD}}$ and the shaded region about the dashed line gives the uncertainty in the predicted torque as given by the spread in $\rho_{\mathrm{TMD}}$. All devices with the possible exception of our monolayer device are well described by the predicted Oersted torque. Deviation from the predicted Oersted torque in our monolayer device may suggest a non-uniform current distribution in the Py film, as cross-sectional HAADF STEM imaging of one of our $\beta$-MoTe$_2$ devices suggests that partial oxidation of the monolayer may play a role in an increased resistivity of that layer, and could affect the growth of the Py on top of such a layer (see Fig.\ \ref{MoTe2Fig5}c and associated discussion).

\section{Dresselhaus-like Torques}\label{MoTe2DressSec}
\begin{figure}[t!]
\centering
\includegraphics[width=7 cm]{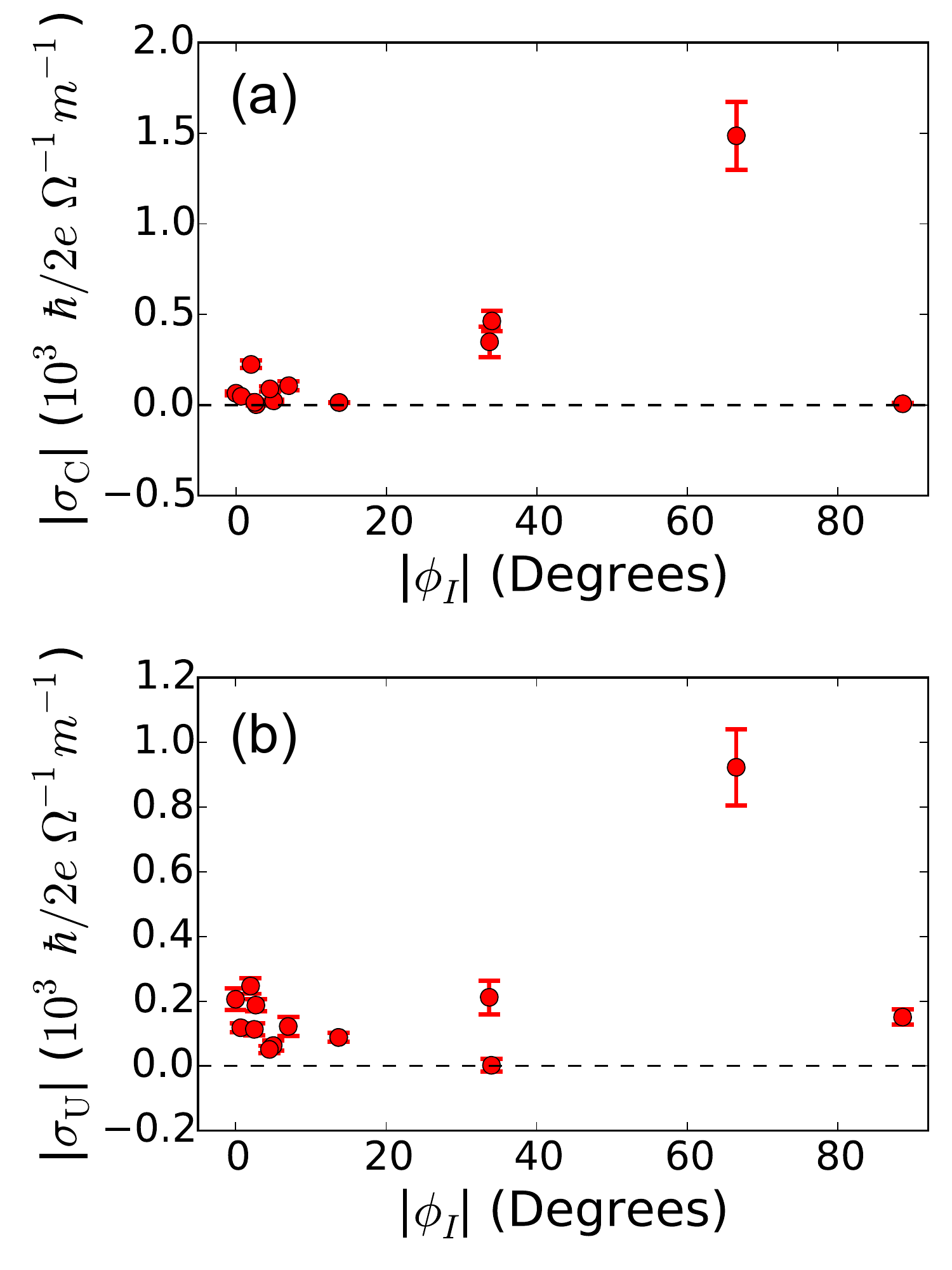}
    \caption{a and b) Torque conductivities for the Dresselhaus-like torques (a) $\propto\hat{m}\times\hat{x}$ and (b) $\propto\hat{m}\times(\hat{m}\times\hat{x})$ as a function of $|\phi_I|$ for all of our ST-FMR devices. The applied microwave power is 5 dBm. Torque conductivities are averaged over measurements at frequencies 8-11 GHz in steps of 1 GHz.}
    \label{MoTe2FigSigmaCandU}
\end{figure}
Figure \ref{MoTe2FigSigmaCandU} shows the torque conductivities for the torque components $\sigma_{\mathrm{C}} \propto\hat{m}\times\hat{x}$, and $\sigma_{\mathrm{U}} \propto\hat{m}\times(\hat{m}\times\hat{x})$ as a function of applied current direction, $\phi_{I}$. (Recall that $\phi_{I}=0^{\mathrm{o}}$ corresponds to current directed perpendicular to the MoTe$_2$ mirror plane.) We refer to the torques $\sigma_{\mathrm{C}}$ and $\sigma_{\mathrm{U}}$ as Dresselhaus-like\ \cite{Stiehl2018}. Symmetry requires that the Dresselhaus-like  torques be zero when the current is either along or perpendicular to a mirror plane ({\it e.g.}$, \phi_{I}=0^{\mathrm{o}}$ or $90^{\mathrm{o}}$, the situations for the majority of the samples studied in this work, including all the samples shown in Fig.\ \ref{MoTe2Fig5}). Consistent with this requirement, in Fig.\ \ref{MoTe2FigSigmaCandU} both $\sigma_{\mathrm{C}}$ and $\sigma_{\mathrm{U}}$ are zero when $\phi_{I}=0^{\mathrm{o}}$ or $90^{\mathrm{o}}$, and can be nonzero at intermediary angles. At present, we do not have enough devices at intermediary values of $\phi_{I}$ to accurately gauge the magnitude of these effects. However, these torque components should arise naturally in MoTe$_2$/ferromagnet heterostructures because MoTe$_2$ has an in-plane resistivity anisotropy\ \cite{HughesJPC1978}, and this will cause  spatially non-uniform current flows with non-zero transverse components whenever the voltage is applied at an angle tilted away from a symmetry axis\ \cite{Stiehl2018}.

\section{In-plane Field-like Torque}\label{mxzappend}

\begin{figure}[t!]
\centering
\includegraphics[width=7 cm]{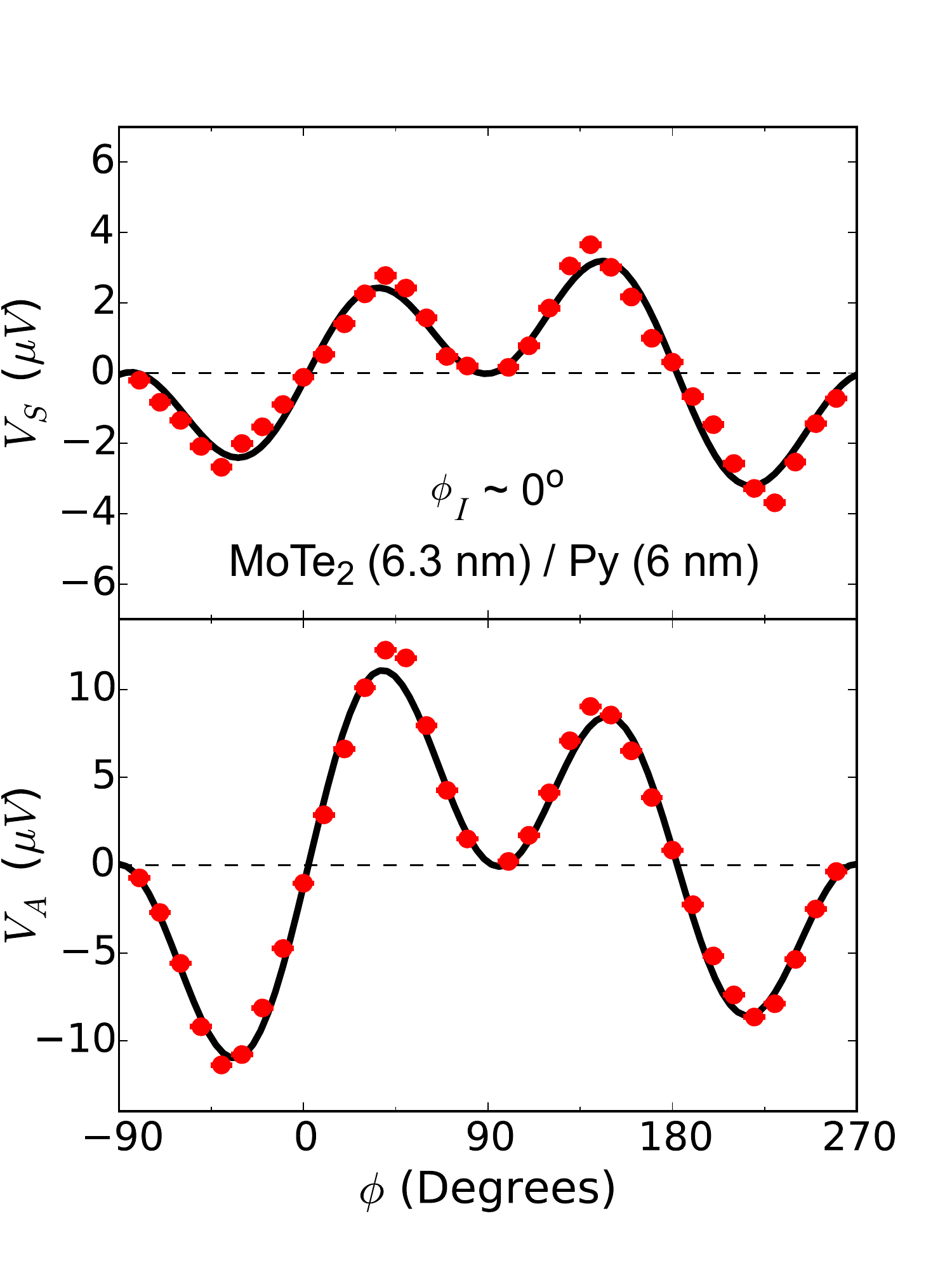}
    \caption{Dependence on the applied magnetic field angle for both the symmetric, $V_{\mathrm{S}}$, and antisymmetric, $V_{\mathrm{A}}$, component of the mixing voltage for a device that shows an in-plane field-like torque $\propto\hat{m}\times\hat{z}$ (Device 3), with MoTe$_2$(6.3 nm) / Py(6 nm) and current applied perpendicular to the MoTe$_2$ mirror plane. Fits of the angular dependence are made using Eqs. \ref{eq:VaAngleMoTe2} and \ref{eq:VsAngleMoTe2}. The applied microwave power is 5 dBm at 9 GHz}
    \label{MoTe2FigSigmaT1}
\end{figure}

\begin{figure}[t!]
\centering
\includegraphics[width=7 cm]{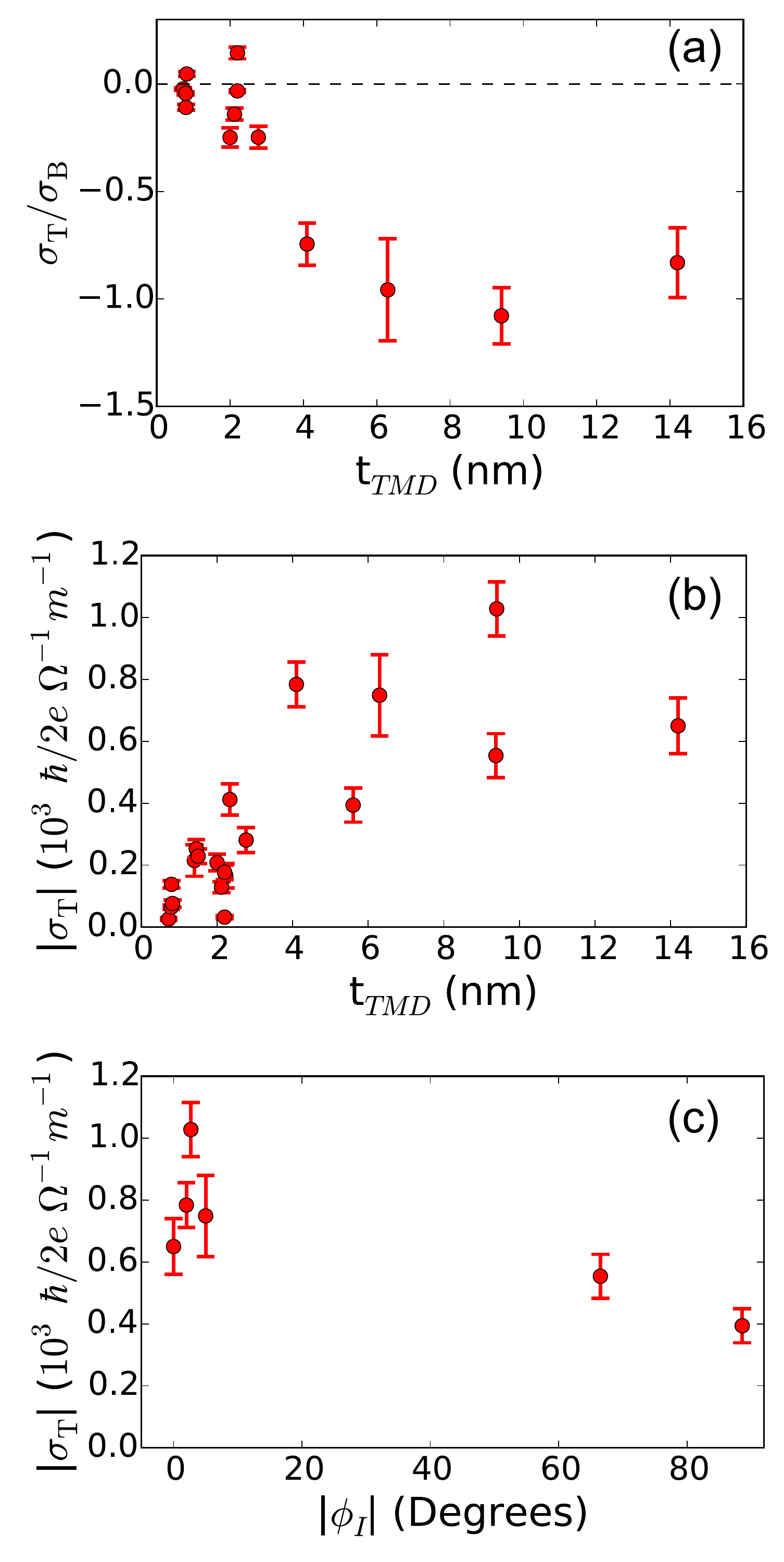}
    \caption{a) The torque ratio $\sigma_{\mathrm{T}}/\sigma_{\mathrm{B}}$ as a function of thickness for devices with the current applied perpendicular to the MoTe$_2$ mirror plane direction. Note that the ratio is always negative when non-zero. We have excluded the bilayer devices as $\sigma_{\mathrm{B}}\sim0$. b) Torque conductivity $|\sigma_{\mathrm{T}}|$ as a function of TMD thickness for all devices. The torque is only appreciable above a TMD thickness of 3 nm. c) $|\sigma_{\mathrm{T}}|$ as a function of $|\phi_{I}|$ for devices with TMD thickness above 3 nm. The applied microwave power is 5 dBm. Torque conductivities are averaged over measurements at frequencies 8-11 GHz in steps of 1 GHz.}
    \label{MoTe2FigSigmaT2}
\end{figure}

The symmetry requirements for the in-plane field-like torque component, $\vec{\tau}_{\mathrm{T}}\propto\hat{m}\times\hat{z}$ are identical to that of $\vec{\tau_{B}}\propto\hat{m}\times(\hat{m}\times\hat{z})$. That is, $\tau_{\mathrm{T}}$ is allowed by symmetry if there is a component of current perpendicular to the single mirror plane. In some but not all of our MoTe$_2$/Py devices we observe a small but nonzero value of $\tau_{\mathrm{T}}$. Figure \ref{MoTe2FigSigmaT1} shows $V_{\mathrm{S}}$ and $V_{\mathrm{A}}$ for one such device (Device 3, with 6.3 nm of MoTe$_2$) for which current is applied perpendicular to the mirror plane. Fitting $V_{\mathrm{A}}$ and $V_{\mathrm{S}}$ with Eqs. \ref{eq:VaAngleMoTe2} and \ref{eq:VsAngleMoTe2} we can extract a ratio of the torques $\tau_{\mathrm{T}}/\tau_{\mathrm{S}}=-0.114\pm0.002$ and $\tau_{\mathrm{T}}/\tau_{\mathrm{B}}=-0.90\pm0.02$.  We observe significant values of $\sigma_{\mathrm{T}}$ only in devices with sufficiently thick MoTe$_2$, above about 3 nm (Fig.\ \ref{MoTe2FigSigmaT2}). The average value of $|\sigma_{\mathrm{T}}|$ for samples with the MoTe$_2$ thickness greater than 3 nm and $\phi_I\approx0^{\mathrm{o}}$ is $810\pm50$ $\hbar/(2e)$ $(\Omega^{-1}m^{-1})$. In all such devices with the current perpendicular to the mirror plane the ratio of $\sigma_{\mathrm{T}}/\sigma_{\mathrm{B}}$ is always negative, even though the signs of $\sigma_{\mathrm{T}}$ and $\sigma_{\mathrm{B}}$ vary from sample to sample (see Fig.\ \ref{MoTe2FigSigmaT2}a). This, together with a similar dependence on sample temperature for $\tau_{\mathrm{T}}$ and $\tau_{\mathrm{B}}$ as discussed in Appendix\ \ref{tempSOTappend}, suggests that these two torque components are correlated. We note, however, that $\sigma_{\mathrm{T}}$ and $\sigma_{\mathrm{B}}$ exhibit very different thickness dependencies, with $\sigma_{\mathrm{T}}$ showing a dependence on thickness typically associated with bulk spin-orbit torque generation, whereas $\sigma_{\mathrm{B}}$ is interfacial in nature. 

Figure\ \ref{MoTe2FigSigmaT2}c shows $|\sigma_{\mathrm{T}}|$ as a function of $|\phi_{I}|$ for devices with TMD thickness above 3 nm. $\sigma_{\mathrm{T}}$ shows a clear decrease in magnitude as the direction of the current is increasingly aligned along the MoTe$_2$ mirror plane. However, near $|\phi_{I}|=90^{\circ}$ there remains a significantly non-zero value of $\sigma_{\mathrm{T}}$ inconsistent with a symmetry analysis of the nominal MoTe$_2$/Py structure. This is reminiscent of the observed $\sigma_{\mathrm{T}}$ in NbSe$_2$/Py devices\ \cite{Guimaraes2018Nano}, in which we presumed  a uniaxial strain induced by the fabrication procedure reduced the nominally high symmetry NbSe$_2$ structure in such a way that this torque could be generated\ \cite{Lee2017}. Note that for (presumably) strained NbSe$_2$/Py devices we observed a large value of $\tau_{\mathrm{T}}$, but no $\tau_{\mathrm{B}}$. 

Together, these observations suggest that there may be two mechanisms that contribution to $\sigma_{\mathrm{T}}$ in $\beta$-MoTe$_2$: one that is correlated with $\sigma_{\mathrm{B}}$ and dependent on the MoTe$_2$ crystal structure, and another that is generated by a symmetry breaking associated with the strain induced during the fabrication procedure.

\section{Determining the Crystal Phase in the Few-Layer Limit}\label{MoTe2Phase}
\begin{figure}[t!]
\centering
\includegraphics[width=7 cm]{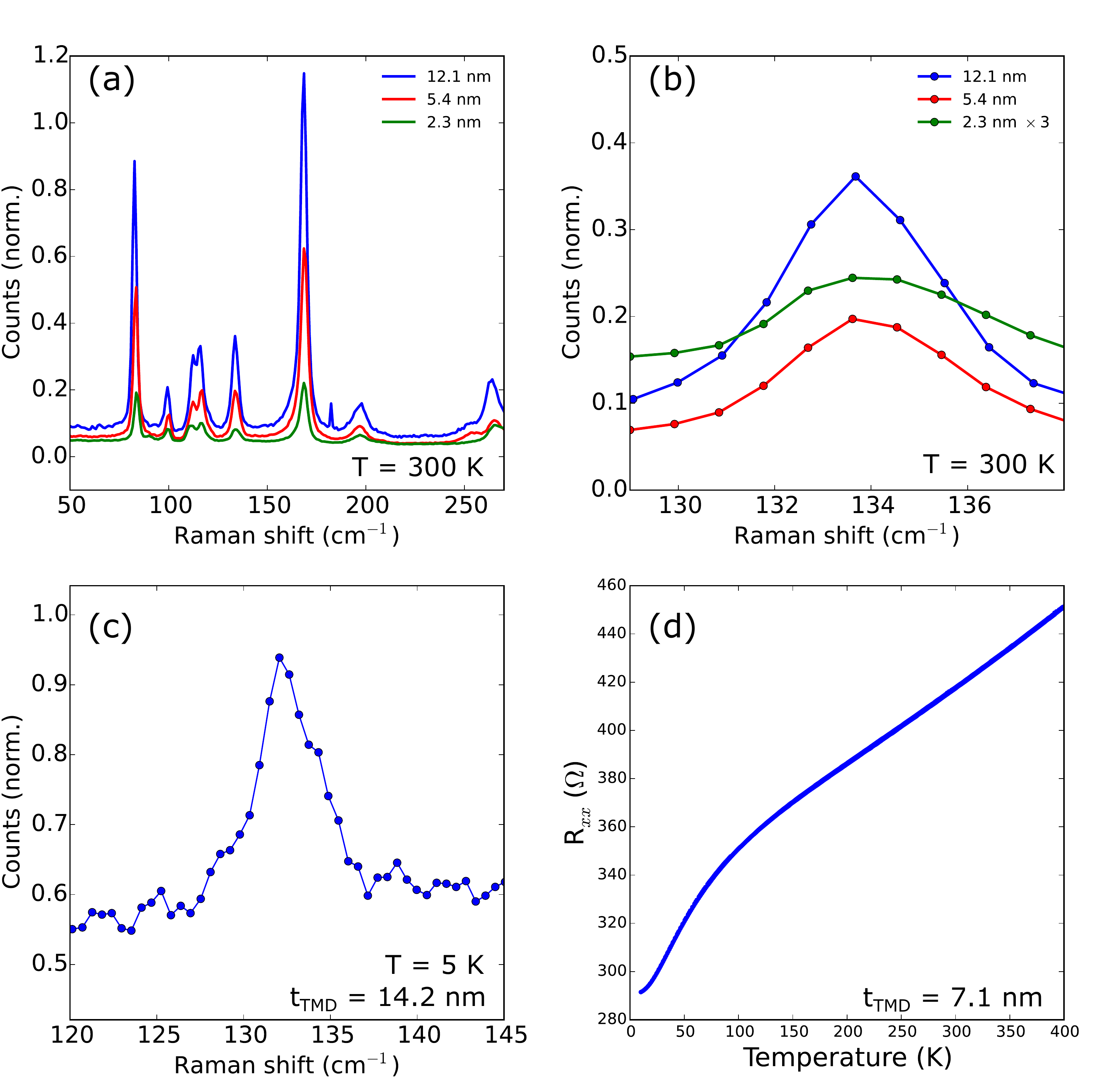}
    \caption{(a) Raman spectroscopy measurements of MoTe$_2$/Py samples with the excitation and collection linearly polarized in a parallel configuration along the Mo chain for three different thicknesses at room temperature. (b) A detailed view of the 133 cm$^{-1}$ mode shown in (a). No peak splitting is observed indicating that the MoTe$_2$ is in the $\beta$ phase. Spectra for (a) and (b) are taken with a 532 nm excitation. (c) Raman spectrum with a 633 nm excitation at 5 K for a 14.2 nm MoTe$_2$/Py device. (d) Four-point resistance of a MoTe$_2$(7.1 nm)/Py device as a function of temperature, with a sample width of 4 $\mu$m, length 15 $\mu$m and a DC current of 50 $\mu$A. }
    \label{MoTe2FigTempDep1}
\end{figure}
\begin{figure}[thb!]
\centering
\includegraphics[width=8 cm]{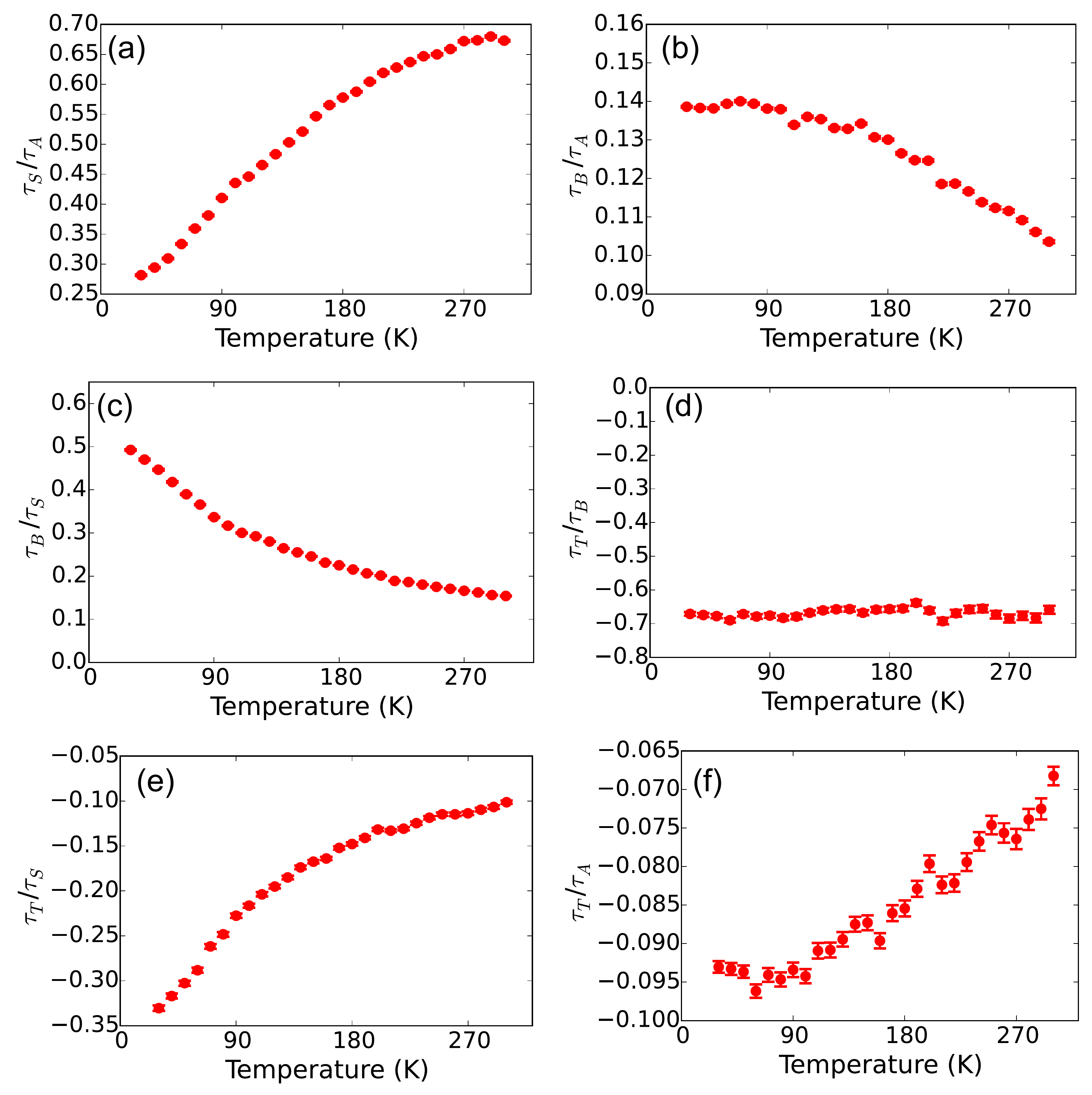}
    \caption{a-e) Torque ratios as a function of temperature for a MoTe$_2$(8.6 nm)/Py device with length 4 $\mu$m and width 3 $\mu$m, with  the current applied perpendicular to the MoTe$_2$ mirror plane ($\phi_I\sim 0^{\circ}$). The applied microwave power is 8 dBm at a frequency of 9 GHz.}
    \label{MoTe2FigTempDep2}
\end{figure}
In bulk crystals, MoTe$_2$ undergoes a hysteretic transition from the $\beta$ phase to the $\gamma$ phase when cooled below approximately 250 K \ \cite{Clarke1978PMB}, with a temperature hysteresis of about 20 K. The orthorhombic ($\gamma$) phase is obtained by a shift in the stacking of the van der Waals layers in $\beta$-MoTe$_2$. This phase is isostructural to WTe$_2$. Both pressure\ \cite{Qi2016,Heikes2018} and impurity doping\ \cite{Rhodes2017,Oliver2D2017,Kim2017PRB} have been shown to influence the transition temperature. While the majority of published studies on $\gamma$-MoTe$_2$ have focused on bulk crystals\ \cite{Clarke1978PMB,Tamai2016PRX,Jiang2017,Deng2016,Huang2016,Zhang2016Raman,ChenNanoLett2016}, a handful of reports have studied the phase transition in thin films\ \cite{He2018PRB,Beams2016ACSnano,ChenNanoLett2016,ZhouJACS2017,ZhongarXiv2018}. Recent work has suggested there may be a thickness dependence to the transition in the few-layer limit\ \cite{He2018PRB,ZhongarXiv2018}. 

The $\beta$ and $\gamma$ phase can be distinguished experimentally through polarized Raman spectroscopy by the presence in the $\gamma$ phase of one additional peak at $\mathtt{\sim}11$ cm$^{-1}$ and a  peak splitting in the $\mathtt{\sim}133$ cm$^{-1}$ mode \cite{Zhang2016Raman,ChenNanoLett2016}. To verify the MoTe$_2$ crystal phase of our devices, we performed Raman spectroscopy measurements of our MoTe$_2$/Py films using 532 nm and 633 nm excitations at room temperature and at 5 K (Fig.\ \ref{MoTe2FigTempDep1}a-c). We see no evidence of the 133 cm$^{-1}$ peak splitting at room temperature for samples with MoTe$_2$ thicknesses from 2.3 nm to 12.1 nm (Fig.\ \ref{MoTe2FigTempDep1}b), indicating that at room temperature our films are in the $\beta$ phase as expected. The 133 cm$^{-1}$ peak splitting in the $\gamma$ phase is approximately 5 cm$^{-1}$ wide, and so should be resolved by these measurements. The Raman measurements at 5 K also do not show a splitting in the 133 cm$^{-1}$ peak (Fig.\ \ref{MoTe2FigTempDep1}c).  Furthermore, measurements of the four-point resistance of a MoTe$_2$/Py device as a function of temperature (Fig.\ \ref{MoTe2FigTempDep1}d) do not show the hysteretic resistivity feature associated with the transition in bulk samples. These data therefore suggest that our thin films are stabilized in the $\beta$ phase, perhaps due to the deposition of the Py, with no transition to the $\gamma$ phase in the measured temperature range.

\section{Temperature Dependent Measurements of Spin-Orbit Torque}\label{tempSOTappend}
One of our original motivations for studying spin-orbit torques generated by MoTe$_2$ was to try to observe changes in the torques as the MoTe$_2$ underwent a phase transition from the $\beta$ to the $\gamma$ phase as a function of decreasing temperature. As noted in the previous section, it turns out that our device structure stabilizes the $\beta$ phase so that we did not observe any transition to the $\gamma$ phase.  Consistent with the lack of a phase transition in the Raman and four-point resistivity data, measurements of the spin-orbit torques as a function of temperature also reveal a smooth evolution, with no indication of an abrupt transition.  

We performed temperature-dependent ST-FMR measurements for one of our MoTe$_2$/Py devices ($t_{\mathrm{TMD}}=8.6$ nm, length 4 $\mu$m and width 3 $\mu$m), where current is applied perpendicular to the MoTe$_2$ mirror plane ($\phi_I\sim 0^{\circ}$). Figure\ \ref{MoTe2FigTempDep2} shows the ratios of the measured torques as a function of temperature, where the torque magnitudes are extracted from the angular dependence of the applied field direction for $V_A$ and $V_S$, as discussed in the main text.  The reason why we plot torque ratios rather than individual values is that it is difficult to calibrate accurately within our cryostat the exact value of the microwave current within the sample. We observe a smooth increase in the Oersted torque $\tau_{\mathrm{A}}$ with decreasing temperature, reflected in a decrease in the ratio $\tau_{\mathrm{S}}/\tau_{\mathrm{A}}$. This is consistent with the decrease in resistivity of MoTe$_2$ as a function of decreasing temperature, but with no indication of a phase transition to the $\gamma$ phase. The out-of-plane antidamping component $\tau_{\mathrm{B}}$  increases as temperature decreases while the corresponding in-plane antidamping component $\tau_{\mathrm{S}}$  decreases, meaning that the total effective tilt angle of the generated antidamping torque is increasingly pulled out of plane with decreasing temperature. The ratio of $\tau_{\mathrm{T}}/\tau_{\mathrm{B}}$ is constant, indicating that the two torques have the same dependence on temperature. This is additional evidence for the conjecture that these two torque components may arise from related microscopic mechanisms (see discussion in Appendix\ \ref{mxzappend}).

%

\end{document}